\documentclass[aps,prl,twocolumn,superscriptaddress,showpacs]{revtex4}
\usepackage{graphicx}
\usepackage{epsfig}
\usepackage{epstopdf}
\usepackage[mathscr]{eucal}
\usepackage{amsmath,amssymb,bm,graphics}
\usepackage[pdftex,letterpaper=true]{hyperref}

\renewcommand{\bar}[1]{\overline{#1}}
\newcommand{\half}{{\frac{1}{2}}}
\newcommand{\mbf}[1]{\mathbf{#1}}

\renewcommand{\bar}[1]{\overline{#1}}
\begin{document}

\def\Dslash{\raise.15ex\hbox{/}\kern-.7em D}
\def\Pslash{\raise.15ex\hbox{/}\kern-.7em P}

\preprint{SLAC-PUB-13422}

\title{Light-Front Holography: A First Approximation to QCD}

\author{Guy F. de T\'eramond}
\affiliation{Universidad de Costa Rica, San Jos\'e, Costa Rica}

\author{Stanley J. Brodsky}
\affiliation{Stanford Linear Accelerator Center, Stanford University,
Stanford, California 94309, USA}

\date{\today}

\begin{abstract}

Starting from the Hamiltonian equation of motion in QCD,
we identify an invariant  light-front coordinate $\zeta$ which
allows the separation of the dynamics of quark and gluon binding
from the kinematics of constituent spin and internal
orbital angular momentum. The result is a single variable light-front 
Schr\"odinger equation for QCD which determines the eigenspectrum
and the light-front wavefunctions of hadrons for general spin and
orbital angular momentum. This light-front wave equation is
equivalent to the
equations of motion which describe the propagation of spin-$J$ modes
on  anti-de Sitter (AdS) space. 

\end{abstract}

\pacs{11.15.Tk, 11.25Tq, 12.38Aw, 12.40Yx}

\maketitle

One of the most important theoretical tools in atomic physics is the
Schr\"odinger equation, which describes the quantum-mechanical
structure of atomic systems at the amplitude level. Light-front
wavefunctions (LFWFs) play a similar role in quantum chromodynamics
(QCD), providing a fundamental description of the structure and
internal dynamics of hadrons in terms of their constituent quarks
and gluons. The light-front wavefunctions of bound states in QCD are
relativistic generalizations of the Schr\"odinger wavefunctions of
atomic physics, but they are determined at fixed light-cone time
$\tau  = t +z/c$ -- the ``front form" introduced by
Dirac~\cite{Dirac:1949cp} -- rather than at fixed ordinary time $t$.
A remarkable feature of LFWFs is the fact that they are frame
independent; i.e., the form of the LFWF is independent of the
hadron's total momentum $P^+ = P^0 + P^3$ and $\vec P_\perp.$

Light-front quantization is the ideal framework to describe the
structure of hadrons in terms of their quark and gluon degrees of
freedom. The simple structure of the light-front vacuum allows an unambiguous
definition of the partonic content of a hadron. Given the LFWFs, one
can compute observables such as hadronic form factors and structure
functions, as well as the generalized parton distributions and
distribution amplitudes which underly hard exclusive reactions. The
constituent spin and orbital angular momentum properties of the
hadrons are also encoded in the LFWFs.

A key step in the analysis of an atomic system such as positronium
is the introduction of the spherical coordinates $r, \theta, \phi$
which  separates the dynamics of Coulomb binding from the
kinematical effects of the quantized orbital angular momentum $L$.
The essential dynamics of the atom is specified by the radial
Schr\"odinger equation whose eigensolutions $\psi_{n,L}(r)$
determine the bound-state wavefunction and eigenspectrum. In this
paper, we show that there is an analogous invariant
light-front coordinate $\zeta$ which allows one to separate the
essential dynamics of quark and gluon binding from the kinematical
physics of constituent spin and internal orbital angular momentum.
The result is a single-variable light-front Schr\"odinger equation for QCD
which determines the eigenspectrum and the light-front wavefunctions
of hadrons for general spin and orbital angular momentum.

Our analysis follows from recent developments in light-front
QCD~\cite{Brodsky:2003px, deTeramond:2005su, Brodsky:2006uqa,
Brodsky:2008pg, Brodsky:2008pf} which have been inspired by the
AdS/CFT correspondence~\cite{Maldacena:1997re}
between string states in anti-de Sitter (AdS) space and conformal
field theories (CFT) in physical space-time. 
The application of AdS space
and conformal methods to QCD can be motivated from the
empirical evidence~\cite{Deur:2008rf} and theoretical 
arguments~\cite{Brodsky:2008be} that the QCD
coupling $\alpha_s(Q^2) $ has an infrared fixed point at low $Q^2.$
The AdS/CFT correspondence has led to insights into the confining
dynamics of QCD and the analytic form of hadronic light-front
wavefunctions. As we have shown recently, there is a remarkable
mapping between the description of hadronic modes in AdS space and
the Hamiltonian formulation of QCD in physical space-time quantized
on the light-front. This procedure allows string modes $\Phi(z)$ in
the AdS holographic variable $z$ to be precisely mapped to the
light-front wave functions  of hadrons in physical space-time in
terms of a specific light-front variable $\zeta$ which measures the
separation of the quark and gluonic constituents within the hadron.
The coordinate $\zeta$ also specifies the light-front (LF) kinetic energy and
invariant mass of constituents. This mapping was originally obtained
by matching the expression for electromagnetic current matrix
elements in AdS space with the corresponding expression for the
current matrix element using light-front theory in physical space
time~\cite{Brodsky:2006uqa}. More recently we have shown that one
obtains the identical holographic mapping using the matrix elements
of the energy-momentum tensor~\cite{Brodsky:2008pf}, thus providing
an important consistency test and verification of holographic
mapping from AdS to physical observables defined on the light front.

The connection between light-front QCD and the description of
hadronic modes on AdS space is physically compelling and
phenomenologically successful. However,  there are lingering
questions in this approach that should be addressed. In particular,
one wants to understand under what approximations (if any) a formal
gauge/gravity correspondence can be established for physical QCD.
This question is most important if QCD is to be described by the low
energy limit of some (yet unknown) string theory in a higher
dimensional space.  In string theory a spin-$J$ hadronic state is
described by a spin-$J$ field, whereas in physical QCD hadrons are
composite and thus are inevitably endowed of orbital angular
momentum. How can this two pictures be compatible? 
The mapping
between string modes in AdS and LFWFs described
in~\cite{Brodsky:2006uqa, Brodsky:2008pf}  is an important step,
but one must also prove that our
identification of orbital angular momentum  is correct and
compatible with the string description in
terms of eigenmodes of total spin $J$. It is also important to
understand the nature and the validity of the approximations
involved in establishing a gauge/gravity correspondence to find a
framework to systematically improve the results.

In this letter we will show that to a first semiclassical approximation,
light-front QCD is formally equivalent to the equations of motion on a fixed AdS$_5$
gravitational background. To prove this, we show that  the LF Hamiltonian
equations of motion of QCD lead to an effective LF equation for
physical modes  $\phi(\zeta)$ which encode the hadronic properties.
This LF equation carries the orbital angular momentum quantum
numbers and is equivalent to 
the propagation of spin-$J$ modes on AdS space.

We express the hadron four-momentum  generator $P =  (P^+, P^-,
\mbf{P}_{\!\perp})$, $P^\pm = P^0 \pm P^3$, in terms of the
dynamical fields, the Dirac field $\psi_+$, $\psi_\pm = \Lambda_\pm
\psi$, $\Lambda_\pm = \gamma^0 \gamma^\pm$, and the transverse field
$\mbf{A}_\perp$ in the $A^+ = 0$ gauge~\cite{Brodsky:1997de}
 \begin{eqnarray} \nonumber
P^-  &\!\!\!=\!&  \half  \! \int \! dx^- d^2 \mbf{x}_\perp  \, \bar \psi_+  \gamma^+
\frac{ \left( i \mbf{\nabla}_{\! \perp} \right)^2 \!\! + m^2 }{ i \partial^+} \psi_+ + {\rm interactions}, \\
\nonumber
P^+ &\!\!\!=\!&   \int \! dx^- d^2 \mbf{x}_\perp \,
 \bar \psi_+ \gamma^+   i \partial^+ \psi_+, \\ \label{eq:P}
\mbf{P}_{\! \perp}  &\!\!\!=\!&  \half \int \! dx^- d^2 \mbf{x}_\perp \,
\bar \psi_+ \gamma^+   i \mbf{\nabla}_{\! \perp} \psi_+,
\end{eqnarray}
where the integrals are over the initial surface $x^+ = 0$, $x^\pm = x^0 \pm x^3$.
The operator $P^-$ generates LF time translations
$\left[\psi_+(x), P^-\right] = i \partial \psi_+(x)/\partial x^+ $,
and the generators $P^+$ and $\mbf{P}_\perp$ are kinematical.
For simplicity we have omitted from (\ref{eq:P})
the contribution from the gluon field $\mbf{A}_\perp$.

The Dirac field operator is expanded as
\begin{multline} \label{eq:psiop}
\psi_+(x^- \!,\mbf{x}_\perp)_\alpha = \sum_\lambda \int_{q^+ > 0} \frac{d q^+}{\sqrt{ 2
 q^+}}
\frac{d^2 \mbf{q}_\perp}{ (2 \pi)^3} \\ \times
\left[b_\lambda (q)
u_\alpha(q,\lambda) e^{-i q \cdot x} + d_\lambda (q)^\dagger
v_\alpha(q,\lambda) e^{i q \cdot x}\right],
\end{multline}
with $u$ and $v$ LF spinors~\cite{Lepage:1980fj}. Similar expansion follows for the
gluon field $\mbf{A}_\perp$.
Using  LF
commutation relations 
$\left\{b(q), b^\dagger(q')\right\}
= (2 \pi)^3 \,\delta (q^+ \! - {q'}^+)
\delta^{(2)} \! \left(\mbf{q}_\perp\! - \mbf{q}'_\perp\right)$,
we find
\begin{equation*} \label{eq:Pm}
P^- \! \! =  \sum_\lambda \! \int \!  \frac{dq^+ d^2 \mbf{q}_\perp}{(2 \pi)^3 }   \,
\!\! \left(\! \frac{ \mbf{q}_\perp^2 \!+ m^2}{q^+} \right) \!
 b_\lambda^\dagger(q) b_\lambda(q) + { \rm interactions},
\end{equation*}
and we recover the LF dispersion relation $q^- = \frac{\mbf{q}_\perp^2 + m^2}{q^+}$, which follows
from the on shell relation $q^2 = m^2$. The LF time evolution operator
$P^-$ is conveniently written as a term which represents the sum of the kinetic energy of all the partons plus a sum of all the interaction terms.

Is is convenient to define a light-front Lorentz invariant Hamiltonian
$H_{LF}= P_\mu P^\mu = P^-P^+  \! - \mbf{P}^2_\perp$ with eigenstates
$\vert \psi_H(P^+, \mbf{P}_\perp, S_z )\rangle$
and eigenmass  $\mathcal{M}_H^2$, the mass spectrum of the color-singlet states
of QCD~\cite{Brodsky:1997de}
\begin{equation} \label{eq:HLF}
H_{LF} \vert \psi_H\rangle = {\cal M}^2_H \vert \psi_H \rangle.
\end{equation}
A state $\vert \psi_H \rangle$ is an expansion 
in multi-particle Fock states
$\vert n \rangle $ of the free LF Hamiltonian:
~$\vert \psi_H \rangle = \sum_n \psi_{n/H} \vert n \rangle$, where
a one parton state is $\vert q \rangle = \sqrt{2 q^+} \,b^\dagger(q) \vert 0 \rangle$.
The Fock components $\psi_{n/H}(x_i, {\mathbf{k}_{\perp i}}, \lambda_i^z)$
are independent of  $P^+$ and $\mbf{P}_{\! \perp}$
and depend only on relative partonic coordinates:
the momentum fraction
 $x_i = k^+_i/P^+$, the transverse momentum  ${\mathbf{k}_{\perp i}}$ and spin
 component $\lambda_i^z$. Momentum conservation requires
 $\sum_{i=1}^n x_i = 1$ and
 $\sum_{i=1}^n \mathbf{k}_{\perp i}=0$.
The LFWFs $\psi_{n/H}$ provide a
{\it frame-independent } representation of a hadron which relates its quark
and gluon degrees of freedom to their asymptotic hadronic state.

We compute $\mathcal{M}^2$ from the hadronic matrix element
$\langle \psi_H(P') \vert H_{LF}\vert\psi_H(P) \rangle  \! = \!
\mathcal{M}_H^2  \langle \psi_H(P' ) \vert\psi_H(P) \rangle$, expanding the initial and final hadronic states in terms of its Fock components. The computation is much simplified in the light-cone
frame $P = \big(P^+, M^2/P^+, \vec{0}_\perp \big)$ where $H_{LF} =  P^+ P^-$.
We find
\vspace{-3pt}
 \begin{multline} \label{eq:M}
 \mathcal{M}_H^2  =  \sum_n  \! \int \! \big[d x_i\big]  \! \left[d^2 \mbf{k}_{\perp i}\right]   
 \sum_q \Big(\frac{\mbf{k}_{\perp q}^2  \! + m_q^2}{x_q} \Big) \\ \times
 \left\vert \psi_{n/H} (x_i, \mbf{k}_{\perp i}) \right \vert^2  + {\rm interactions} ,
 \end{multline}
plus similar terms for antiquarks and gluons ($m_g = 0)$. The integrals in (\ref{eq:M}) are over
the internal coordinates of the $n$ constituents for each Fock state with phase space normalization
\begin{equation}
\sum_n  \int \big[d x_i\big] \left[d^2 \mbf{k}_{\perp i}\right]
\,\left\vert \psi_{n/H}(x_i, \mbf{k}_{\perp i}) \right\vert^2 = 1.
\end{equation}
The LFWF $\psi_n(x_i, \mathbf{k}_{\perp i})$ can be expanded in terms of  $n-1$ independent
position coordinates $\mathbf{b}_{\perp j}$,  $j = 1,2,\dots,n-1$, so that ~$\sum_{i = 1}^n \mbf{b}_{\perp i} = 0$. We can also express (\ref{eq:M})
in terms of the internal coordinates $\mbf{b}_{\perp j}$ with $\mbf{k}_\perp^2  \to
- \nabla_{\mbf{b}_\perp}^2$.
The normalization is defined by
\vspace{-4pt}
\begin{equation}  \label{eq:Normb}
\sum_n  \prod_{j=1}^{n-1} \int d x_j d^2 \mathbf{b}_{\perp j}
\vert \psi_{n/H}(x_j, \mathbf{b}_{\perp j})\vert^2 = 1.
\end{equation}

To simplify the discussion we will consider a two-parton hadronic bound state. In the limit
$m_q \to 0$
\begin{eqnarray} \nonumber
\mathcal{M}^2  &\!\!=\!\!&  \int_0^1 \! d x \! \int \!  \frac{d^2 \mbf{k}_\perp}{16 \pi^3}   \,
  \frac{\mbf{k}_\perp^2}{x(1-x)}
 \left\vert \psi (x, \mbf{k}_\perp) \right \vert^2  + {\rm interactions} \\ \nonumber
  &\!\!=\!\!& \int_0^1 \! \frac{d x}{x(1-x)} \int  \! d^2 \mbf{b}_\perp  \,
  \psi^*(x, \mbf{b}_\perp)
  \left( - \mbf{\nabla}_{ {\mbf{b}}_\perp}^2\right)
  \psi(x, \mbf{b}_\perp) \\   \label{eq:Mb} & & ~~~~~~~~ +  {\rm interactions}.
 \end{eqnarray}

 It is clear from (\ref{eq:Mb}) that the functional dependence  for a given Fock state is
given in terms of the invariant mass
\begin{equation}
 \mathcal{M}_n^2  = \Big( \sum_{a=1}^n k_a^\mu\Big)^2 = \sum_a \frac{\mbf{k}_{\perp a}^2 \! + m_a^2}{x_a}
 \to \frac{\mbf{k}_\perp^2}{x(1-x)} \,,
 \end{equation}
 the measure of the off-mass shell energy~ $\mathcal{M}^2 - \mathcal{M}_n^2$.
 Similarly in impact space the relevant variable for a two-parton state is  $\zeta^2= x(1-x)\mbf{b}_\perp^2$.
Thus, to first approximation  LF dynamics  depend only on the boost invariant variable
$\mathcal{M}_n$ or $\zeta$
and hadronic properties are encoded in the hadronic mode $\phi(\zeta)$:
 $ \psi(x, \mbf{k}_\perp) \to \phi(\zeta)$.
 We choose the normalization of  the LF mode $\phi(\zeta) = \langle \zeta \vert \phi \rangle$
 \begin{equation}
 \langle\phi\vert\phi\rangle = \int \! d \zeta \,
 \vert \langle \zeta \vert \phi\rangle\vert^2 = 1.
 \end{equation}

We write the Laplacian operator in (\ref{eq:Mb}) in circular cylindrical coordinates
$(\zeta, \varphi)$ with $\zeta = \sqrt{x(1-x)} \, \vert\mbf{b}_\perp\vert$:
$\nabla_\zeta^2 = \frac{1}{\zeta} \frac{d}{d\zeta} \left( \zeta \frac{d}{d\zeta} \right)
+ \frac{1}{\zeta^2} \frac{\partial^2}{\partial \varphi^2}$, and factor out the angular dependence of the
modes in terms of the $SO(2)$ Casimir representation $L^2$ of orbital angular momentum in the
transverse plane:
$\phi(\zeta, \varphi) \sim e^{\pm i L \varphi} \phi(\zeta)$. Expressing the LFWF $ \psi(x, \zeta)$ as a product
of the LF mode $\phi(\zeta)$ and a prefactor $f(x)$
\begin{equation} \label{eq:psiphi}
\psi(x,\zeta) = \frac{\phi(\zeta)}{\sqrt{2 \pi \zeta}} f(x).
\end{equation}
we find
\begin{eqnarray} \nonumber
\mathcal{M}^2  &\!\!=\!\!& \int \! d\zeta \, \phi^*(\zeta) \sqrt{\zeta}
\left( -\frac{d^2}{d\zeta^2} -\frac{1}{\zeta} \frac{d}{d\zeta}
+ \frac{L^2}{\zeta^2}\right)
\frac{\phi(\zeta)}{\sqrt{\zeta}}  \\
&&   ~~~~~~~~~~~~~~~~~
 + \int \! d\zeta \, \phi^*(\zeta) U(\zeta) \phi(\zeta) \\ \nonumber
&\!\!=\!\!&
\int \! d\zeta \, \phi^*(\zeta)
\left( -\frac{d^2}{d\zeta^2}
- \frac{1 - 4L^2}{4\zeta^2} +U(\zeta)\right)
\phi(\zeta),
\end{eqnarray}
where  the complexity of the interaction terms in the QCD Lagrangian is summed up in the addition of the effective
 potential $U(\zeta)$, which is then modeled to enforce confinement at some IR scale. 
The light-front eigenvalue equation $H_{LF} \vert \phi \rangle = \mathcal{M}^2 \vert \phi \rangle$
is thus a light-front wave equation for $\phi$
\begin{equation} \label{eq:QCDLFWE}
\left(-\frac{d^2}{d\zeta^2}
- \frac{1 - 4L^2}{4\zeta^2} + U(\zeta) \right)
\phi(\zeta) = \mathcal{M}^2 \phi(\zeta),
\end{equation}
an effective single-variable light-front Schr\"odinger equation which is
relativistic, covariant and analytically tractable. From (\ref{eq:M}) one can readily
generalize the equations to allow for the kinetic energy of massive
quarks~\cite{Brodsky:2008pg}.

As the simplest example we consider a bag-like model~\cite{Chodos:1974je}
where  partons are free inside the hadron
and the interaction terms effectively build confinement. The effective potential is a hard wall:
$U(\zeta) = 0$ if  $\zeta \le 1/\Lambda_{\rm QCD}$ and
 $U(\zeta) = \infty$ if $\zeta > 1/\Lambda_{\rm QCD}$.
 However, unlike the standard bag model~\cite{Chodos:1974je}, boundary conditions are imposed on the
 boost invariant variable $\zeta$, not on the bag radius at fixed time.
 If $L^2 \ge 0$ the LF Hamiltonian is positive definite
 $\langle \phi \vert H_{LF} \vert \phi \rangle \ge 0$ and thus $\mathcal M^2 \ge 0$.
 If $L^2 < 0$ the LF Hamiltonian is unbounded from below and the particle
 ``falls towards the center''~\cite{LL:1958}. The critical value corresponds to $L=0$.
  The mode spectrum  follows from the boundary conditions
 $\phi \! \left(\zeta = 1/\Lambda_{\rm QCD}\right) = 0$, and is given in
 terms of the roots of Bessel functions: $\mathcal{M}_{L,k} = \beta_{L, k} \Lambda_{\rm QCD}$.
 Since in the conformal limit $U(\zeta) \to 0$, Eq. (\ref{eq:QCDLFWE}) is equivalent to an AdS
 wave equation. The hard-wall LF model discussed here is equivalent to the hard wall model of
 Ref.~\cite{Polchinski:2001tt}. Likewise a two-dimensional  oscillator with
 effective potential $U(\zeta) \sim \zeta^2$ is equivalent to the soft-wall model of
 Ref.~\cite{Karch:2006pv} which reproduce the usual linear Regge trajectories.

 We are now in a position to find out if the first order approximation to light-front QCD discussed above
admits an effective gravity description. 
To examine this question it is useful to study the structure of the equation of motion of
$p$-forms in AdS space,
which for $p=0$ and $p = 1$ represent spin 0 and spin 1 states respectively.
A $p$-form in AdS is
a totally antisymmetric tensor field
$\Phi_{\ell_1 \ell_2 \cdots \ell_p}$ of rank $p$
which couples to an interpolating operator $\mathcal{O}$ of
dimension $d-p$ at the AdS boundary. Fermionic modes will be described elsewhere.
In tensor notation the equations of motion for a p-form are expressed as the set of $p+1$
coupled equations~\cite{l'Yi:1998eu}
\begin{multline} 
\big[ z^2 \partial_z^2 - (d + 1 - 2 p) z \, \partial_z - z^2 \partial_\rho \partial^\rho  \\
    ~~~~~~~~~~~~~~~~~~~~~~ - (\mu R)^2 +  d + 1 - 2 p \big] \Phi_{z  \alpha_2 \cdots \alpha_p}  
      = 0, \\
     \cdots \\ 
\left[ z^2 \partial_z^2 - (d - 1 -2 p) z \, \partial_z - z^2 \partial_\rho \partial^\rho
     - (\mu R)^2  \right]  \Phi_{\alpha_1  \alpha_2 \cdots \alpha_p} \\
    = 2z \bigl( \partial_{\alpha_1}  \Phi_{z  \alpha_2 \cdots \alpha_p}
     + \partial_{\alpha_2} \Phi_{\alpha_1  z \cdots \alpha_p}
     + \cdots\bigr), \label{eq:eomPhi3}
\end{multline}
where $\mu$ is a $d\!+\!1$-dimensional mass, $\rho = 0,1, \cdots, d\!-\!1$ and $R$ is the AdS$_{d+1}$ radius. 

Consider the plane-wave solution $\Phi_P(x,z) _{\alpha_1 \cdots \alpha_p}  \! = \! 
e^{- i P \cdot x} \, \Phi(z)_{\alpha_1  \cdots \alpha_p}$, with four-momentum
$P_\mu$, invariant hadronic mass $P_\mu P^\mu = \mathcal{M}^2$ and spin
indices $\alpha$ along the 3+1 physical coordinates. For string modes with all  indices 
along the Poincar\'e coordinates,
$\Phi_{z  \alpha_2 \cdots \alpha_p} = \Phi_{\alpha_1  z \cdots \alpha_p} = \cdots = 0$,
the coupled differential  equations  (\ref{eq:eomPhi3}) reduce to the  homogeneous wave equation
\begin{equation} \label{eq:eomPhipz}
\left[ z^2 \partial_z^2 - (d\! -\! 1 \!- \!2 p) z \, \partial_z + z^2 \mathcal{M}^2
\!  -  (\mu R)^2 \right] \!  \Phi _{\alpha_1  \cdots \alpha_p}  = 0,
\end{equation}
with conformal dimension
\begin{equation}
\Delta = \frac{1}{2} \bigl( d \! + \! \sqrt{ (d - 2 p)^2 + 4 \mu^2 R^2} \bigr).
\end{equation}
and thus   $(\mu R)^2 = (\Delta-p)(\Delta-d+p)$.

Thus when the  polarization indices are chosen along the physical $ 3+1$ Poincar\'e  
coordinates, the p-form equation  (\ref{eq:eomPhipz}) becomes homogeneous and its 
polarization structure decouples; i.e., it is independent of the kinematical polarization structure of the indices.  
Thus  it also 
describes the dynamics of a spin $J\!=\!p$-mode in AdS $\Phi(x,z) _{\mu_1 \cdots \mu_J}$, which is 
totally symmetric in all its indices. To prove this, consider  the AdS wave equation (\ref{eq:eomPhipz}) for
a scalar mode $\Phi$ ($p=0$),
and define  a  spin-$J$ field $\Phi_{\mu_1 \cdots \mu_J}$ with shifted dimensions:
$\Phi_J(z) =  \left(z/R\right)^{-J}  \Phi(z)$,
and normalization~\cite{Hong:2004sa}
\begin{equation} 
R^{d-2J-1} \int_0^{z_{max}} \! \frac{dz}{z^{d-2J-1}} \, \Phi_J^2 (z) = 1.
\end{equation}
The shifted field $\Phi_J$ obeys the equation of motion
\begin{equation} \label{eq:eomPhiJz}
\left[ z^2 \partial_z^2 - (d\! -\! 1 \!- \!2 J) z \, \partial_z + z^2 \mathcal{M}^2
\!  -  (\mu R)^2 \right] \!  \Phi_J  = 0,
\end{equation}
where the fifth dimensional mass is rescaled according to $(\mu R)^2 \to (\mu R)^2 - J(d-J)$.
One can then construct an effective action in terms of high spin modes
$\Phi(x,z) _{\mu_1 \cdots \mu_J}$ with only the physical degrees of 
freedom~\cite{Karch:2006pv}.

Upon the substitution $z \! \to\! \zeta$  and
$\phi_J(\zeta)   \!  \sim \! \zeta^{-3/2 + J} \Phi_J(\zeta)$
in (\ref{eq:eomPhiJz})
we recover for $d=4$ the QCD light-front  wave equation (\ref{eq:QCDLFWE})
in the conformal limit
\begin{equation} \label{eq:ScheqS}
\left(-\frac{d^2}{d \zeta^2} - \frac{1-4 L^2}{4\zeta^2} \right) \phi_{\mu_1 \cdots \mu_J}
= \mathcal{M}^2 \phi_{\mu_1 \cdots \mu_J},
\end{equation}
where the fifth dimensional mass is not a free parameter but scales according to
$ (\mu R)^2 = - (2-J)^2 + L^2$.
In the hard-wall model there is a total decoupling of the total spin $J$.
For $L^2 \ge 0$  the LF Hamiltonian is positive definite
$\langle \phi_J \vert H_{LF} \vert \phi_J \rangle \ge 0$
and we find the stability bound $(\mu R)^2 \ge -  (2 - J)^2$.
For $J = 0$ the stability condition gives the bound
$(\mu R)^2 \ge - 4$. The quantum-mechanical stability conditions discussed here are thus
equivalent to the Breithelohner-Freedman stability bound in AdS~\cite{Breitenlohner:1982jf}.
The scaling dimensions are $\Delta = 2 + L$ independent of $J$ in agreement with the
twist scaling dimension of a two parton bound state in QCD.

We have shown that the use of the invariant coordinate $\zeta$ in light-front QCD which is related to the
fundamental constituent structure,
allows the separation of the dynamics of quark and gluon binding
from the kinematics of constituent spin and internal
orbital angular momentum. The result is a single-variable LF
Schr\"odinger equation  which determines the spectrum
and  LFWFs of hadrons for general spin and
orbital angular momentum. 
This LF wave equation serves as a semiclassical first approximation to QCD and is equivalent to the
equations of motion which describe the propagation of spin-$J$ modes
on  AdS. 
Remarkably, the AdS equations
correspond to the kinetic energy terms of  the partons inside a
hadron, whereas the interaction terms build confinement and
correspond to the truncation of AdS space.
As in this approximation there are no quantum corrections, there are no
anomalous dimensions. This may explain the experimental success of
power-law scaling in hard exclusive reactions where there are no
indications of  the effects of anomalous dimensions. 
In the hard-wall model 
there is total orbital decoupling from hadronic spin
$J$ and thus the LF excitation spectrum
of hadrons  depends only on 
orbital and principal quantum numbers. In this
model the mass dependence has the linear form:  $\mathcal{M} \sim 2n + L$. In the soft-wall
model the usual Regge behavior is found $\mathcal{M}^2 \sim n +
L$ where the slope in $L$ and $n$ is identical.  Both models predict the same multiplicity of states for mesons
and baryons as observed experimentally~\cite{Klempt:2007cp}.
As in the Schr\"odinger equation, the semiclassical approximation to light-front QCD 
described in this letter does not account for particle
creation and absorption, and thus it is expected to break down at short distances
where hard gluon exchange and quantum corrections become important. However, 
one can systematically improve the holographic approximation by
diagonalizing the QCD light-front Hamiltonian on the AdS/QCD basis.

\begin{acknowledgments}

We thank Andreas Karch, Igor Klebanov, Leonardo Rastelli, Robert
Shrock, Matt Strassler and James Vary for helpful conversations. This research
was supported by the Department of Energy contract
DE--AC02--76SF00515.

\end{acknowledgments}

\end{document}